\documentclass[twocolumn,aps,10pt,prb,showpacs,showkeys,]{revtex4}
\usepackage[dvips]{graphicx}

\begin{document}
\title{Electronic structure and magnetism of a ferromagnetic insulator Cs$_{2}$AgF$_{4}$:\\
explanation for a lilac colour and a prediction for the
anisotropic $g$ factor}
\author{R. J. Radwanski}
\affiliation{Center of Solid State Physics, S$^{nt}$Filip 5,
31-150 Krakow, Poland,
\\
Institute of Physics, Pedagogical University, 30-084 Krakow,
Poland} \homepage{http://www.css-physics.pl}
\email{sfradwan@cyf-kr.edu.pl}
\author{Z. Ropka}
\affiliation{Center of Solid State Physics, S$^{nt}$Filip 5,
31-150 Krakow, Poland}

\begin{abstract}
Magnetic properties of stechiometric Cs$_{2}$AgF$_{4}$ have been
calculated within a very strong correlation limit taking into
account a low-symmetry crystal field and the intra-atomic
spin-orbit coupling of the Ag$^{2+}$ ion. We consistently explain
the insulating ground state and the magnetic state revealing the
spin gap of 2.6 meV below T$_{c}$ of 14.9 K. A $d-d$
excitation of 2.0-2.3 eV related to the $t_{2g}$-$e_{g}$ promotion
energy (=10$Dq$) is a reason for the lilac colour of
Cs$_{2}$AgF$_{4}$. Our approach can be experimentally verified by
the measurement of the $g$ factor ($g_{z}$ =2.12 and $g_{y}$
=2.52) and the absorption energy at 2.0-2.3 eV.

\end{abstract}
\maketitle

\vspace {-0.6 cm} \section {Introduction}

Cs$_{2}$AgF$_4$ is a unique 4$d$ ferromagnet \cite{1,2} - the
most of fluorides and oxides are antiferromagnetic. Despite that
Cs$_{2}$AgF$_4$ has already been synthesized 30 years ago
\cite{2} recently it draws attention \cite{3,4,5,6,7} being
regarded as an analog of La$_{2}$CuO$_4$, a maternal
high-temperature superconductor. Thus it seems to be very good
examplary system for studying basic interactions in $d$ fluorides
and oxides.

Cs$_{2}$AgF$_{4}$, when stoichiometric, is a good insulator. It
is ferromagnetic below T$_{c}$ of 14.9 K \cite{1} or 13.95 K
\cite{3}. The macroscopic magnetisation, if recalculated per the
formula unit, points to a moment of about 0.8 $\mu_{B}$ \cite{1}.
The paramagnetic susceptibility has been found to follow the
Curie-Weiss law \cite{1}. Already above 50 K there is a straight
$\chi^{-1}$ {\it vs} T line with $\theta_{CW}$ = +30 K.

The origin of the ferromagnetic state and the insulating state is
recently of a large theoretical discussion in Phys. Rev. journals
\cite{3,4,5,6,7,8}. As far as the ferromagnetic state is discussed
another fluoride ferromagnetic compound K$_{2}$CuF$_{4}$ has been
recalled. They both have similar structure based on the
K$_{2}$NiF$_{4}$ structure. Kasinathan {\it et al.} \cite{4} have
explained, within density-functional theory (DFT), the
ferromagnetic structure as originating from the substantial Ag-F
covalency. But this covalency causes simultaneously the incorrect
itinerant, not insulating, ground state and a substantial
magnetic moment on the fluorine ions. Dai {\it et al.} \cite{5}
suggested that the ferromagnetism originates from the spin
polarization induced by the d$_{z^{2}}$-p-d$_{x^{2}-y^{2}}$
orbital interaction through the Ag-F-Ag bridges. In Refs
\cite{4,5} authors dealt with the tetragonal structure and
obtained a half-metallic solution in the ferromagnetic state -
this half-metallic solution, being essential for their explanation
of the ferromagnetism in Cs$_{2}$AgF$_{4}$, disagrees with the
insulating ground state observed experimentally. In following
studies Kan {\it et al.} \cite {6}, performing pseudopotential
DFT calculations for the orthorhombic lattice, have obtained an
orbitally ordered solution but with unphysically large in-plane
and out-of-plane magnetic coupling strengths. More recently, Hao
{\it et al.} \cite{7} and Wu and Khomskii \cite{8}, using DFT
total-energy calculations, have found that the orthorhombic
lattice is more energetically stable than the undistorted
tetragonal lattice. Moreover they theoretically found that this
inherent lattice distortion is accompanied by the Ag 4$d$-orbital
ordering and this orbital ordering accounts for the observed
ferromagnetism of Cs$_{2}$AgF$_{4}$ similarly like for an
isoelectronic and isostructural compound, K$_{2}$CuF$_{4}$. Wu
and Khomskii found, within the GGA+U calculations with U = 3 eV,
that "Cs$_{2}$AgF$_{4}$ is stabilized in an insulating
orthorhombic phase rather than in a metallic tetragonal phase"
and that "the ground state is orbitally ordered ferromagnetic
state". This orbitally ordered ferromagnetic ground state is
realized by the alternative hole occupation in the
$x^{2}-z^{2}$/$y^{2}-z^{2}$ orbitals.

Analyzing this recent theoretical work of Wu and Khomskii we
would like to put attention to the following outcomes:

1) the orthorhombic distortion is necessary to get insulating
ground state - within the GGA approach with U=0 there is a small gap of
0.2 eV only; this gap increases to 1 eV for an expected U value of 3 eV;

2) the insulating gap of 0.2 eV, as well as 1 eV, is formed within
{\bf the spin-polarized} $x^{2}-z^{2}$ state (Fig. 4 \cite{8});

3) a size of the spin splitting is very large - in case of the
$x^{2}-z^{2}$ state the spin splitting amounts to 0.75 eV (U=0,
see Fig. 1) and it increases to 1.5 eV for the final calculations
with U=3 eV (Fig. 4 \cite{8}).

4) there is a relatively strong spin polarization of the F atoms
- the local spin moment of each apical F (F$_{1}$) amounts to
0.099 $\mu_{B}$ and of each planar F (F$_{2}$) amounts to 0.097
$\mu_{B}$. Thus all fluorine atoms contribute by 0.392 $\mu_{B}$
per formulae unit to the resultant magnetisation. It is 40 \% of
the calculated total magnetisation, of 0.992 $\mu_{B}$/f.u..

5) there is a relatively small magnetic moment of Ag, of 0.600
$\mu_{B}$ only; thanks the large F-atom contribution the
resultant magnetisation, being 0.992 $\mu_{B}$, becomes close to
the integer 1 $\mu_{B}$, expected for a S = 1/2 spin.

The aim of this paper is to present results of calculations of
properties of Cs$_{2}$AgF$_{4}$ within the very strong-correlation
limit.

\section {Theoretical outline}
The very strong-correlation limit has a lot in common with the ionic
model and the many-electron crystal-field model which we consider
to be to a large extent relevant to the reality of oxides and
fluorides. We would like to note that we employ the many-electron
version of the crystal-field theory instead of the one-electron
version mentioned in Refs \cite{4,5,6,7,8}. Our electronic
structure is different from those obtained in Refs
\cite{4,5,6,7,8} despite that energies of $e_{g}$ and $t_{2g}$
states are discussed.
\begin{figure}[t]
\begin{center}
\includegraphics[width = 8.1 cm]{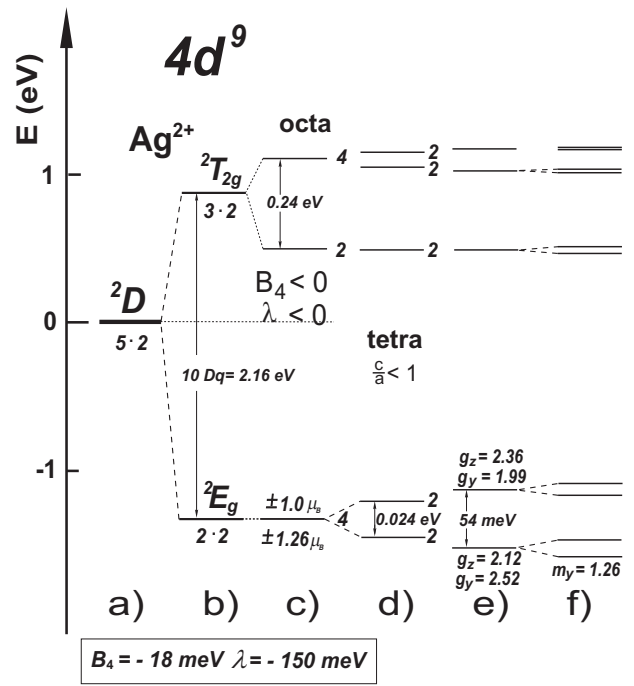}
\end{center} \vspace {-0.2 cm}
\caption{The calculated fine electronic structure of the 4$
d^{9}$ electronic system (Ag$^{2+}$, Cu$^{2+}$(3$ d^{9}$) ions)
in the paramagnetic state under the action of the crystal field
and spin-orbit interactions: a) the 10-fold degenerated $^2$D term
realized in the absence of the CEF and the s-o interactions; b)
the splitting of the $^2$D term by the octahedral CEF surrounding B$_4$%
= -18 meV ($\lambda _{s-o}$ =0) yielding the $^2$E$_{g}$ cubic
subterm as the ground state and 10Dq = 2.16 eV; c) the effect of
the spin-orbit ($\lambda _{s-o}$ = -150 meV) for the octahedral
CEF states causing a splitting of the higher $^2$T$_{2g}$ cubic
subterm; the degeneracy and the associated magnetic moments are
shown; d) the splitting due to the compressed tetragonal
off-octahedral distortion of B$_2^0$= +2 meV (c/a$<$1 apical
fluorines become closer); e) the splitting due to the in-plane
off-tetragonal distortion (elongation along y-axis) of B$_2^2$=
-7 meV); f) - the splitting in the magnetic state. Figs c, d, e and f
are not to the left hand energy scale. \vspace {-0.2 cm}}
\end{figure}
According to us, and in contrary to the above-mentioned papers
\cite{4,5,6,7,8}, strong correlations are realized i) by the
relevant charge transfer during the formation of the compound
which leads to the charge distribution
Cs$_{2}^{1+}$Ag$^{2+}$F$_{4}^{1-}$ and ii) by strong
correlations among nine $d$ electrons of the Ag$^{2+}$ ion. These
strong intra-atomic correlations among nine $d$ electrons assure
that they should be considered as the whole 4$d^{9}$ system being
described by quantum numbers L=2 and S=1/2 (term $^{2}D$). In
Cs$_{2}$AgF$_{4}$ the octahedral fluorine surroundings splits 10
states for the lowest four ($^{2}E_{g}$) and higher six states
($^{2}T_{2g}$) (Fig. 1b). The compressed tetragonal distortion
occurring in Cs$_{2}$AgF$_{4}$ causes a splitting of the lowest
quartet for two Kramers doublets (Fig. 1d). The longest bond
along y-axis occurring in Cs$_{2}$AgF$_{4}$ enlarges only this
splitting (Fig. 1e) as the Kramers doublet cannot be split by any
lattice distortion. The Kramers doublet is only split by a
magnetic field, external or internal. We have calculated that the
Ag$^{2+}$ moment experiences internal molecular field of 17.4 T
(at T= 0 K).

\section {Results and discussion}
We accept experimental lattice parameters at T = 293 K, according
to Ref. \cite{1}, cited by \cite {8}: $a_{o}$ =643.5 pm,
$b_{o}$=643.9 pm and $c_{o}$ = 14.150 pm and the respective Ag-F
lengths: in-plane 216.8 pm (Ag-F$_{2}$(a)), 238.3 pm
(Ag-F$_{2}$(b)) and the apical bond 217.2 pm (Ag-F$_{1}$(c)).

\begin{figure}[t]
\begin{center}
\includegraphics[width = 7.1 cm]{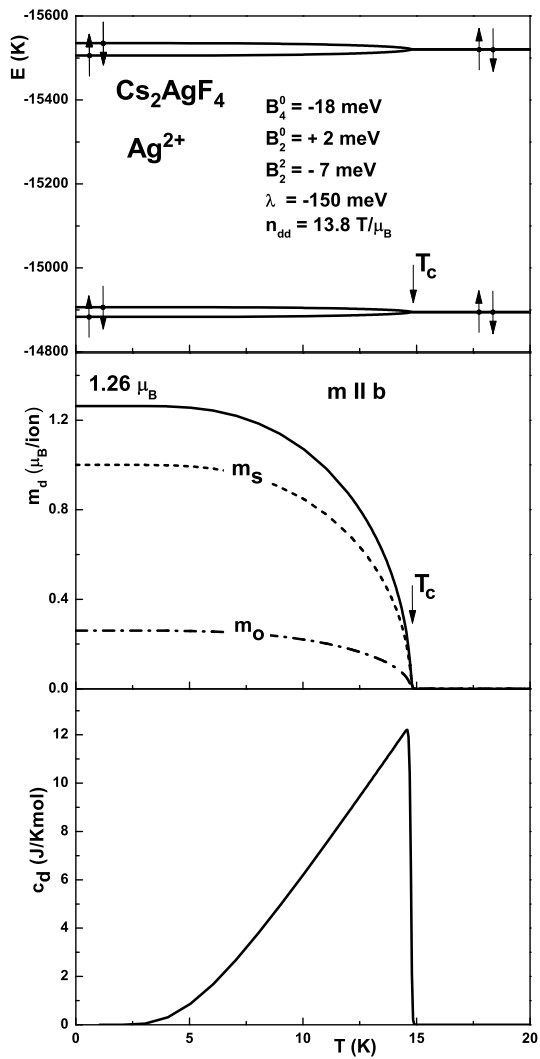}
\end{center} \vspace {-0.3 cm}
\caption{The calculated temperature dependence of some properties
of Cs$_{2}$AgF$_{4}$. a) the temperature dependence of the two
lowest states ($^{2}E_{g}$ states) of the Ag$^{2+}$-ion in
Cs$_{2}$AgF$_{4}$ in the magnetically-ordered state below T$_{c}$
of 14.9 K; in the paramagnetic state the electronic structure is
temperature independent unless we do consider a changing of the
CEF parameters, for instance, due to the thermal lattice
expansion. The used parameters: B$_4$= -18 meV, B$_2^{0}$= +2
meV, B$_2^{2}$= -7 meV, $\lambda _{s-o}$ = -150 meV and
n$_{d-d}$= 13.8 T/$\mu_{B}$. In the ferromagnetic state the
Kramers doublets become split. Excited states are at 2.06, 2.28
and 2.30 eV. (b) the temperature dependence of the Ag$^{2+}$-ion
magnetic moment in Cs$_{2}$AgF$_{4}$. At 0 K the total moment
m$_{Ag}$ of 1.26 $\mu _{B}$ is built up from the orbital m$_{o}$
and spin m$_{s}$ moment of 0.26 and 1.00 $\mu _{B}$,
respectively. c) The calculated temperature dependence of the
4$d$ contribution $c_{d}(T)$ to the heat capacity of
Cs$_{2}$AgF$_{4}$. The $\lambda$-type peak marks T$_{c}$. \vspace
{-0.5 cm}}
\end{figure}
This situation in Cs$_{2}$AgF$_{4}$ we account for by
crystal-field parameters (z along the c axis) of
Cs$_{2}$AgF$_{4}$: B$_4$ = -18 meV (octupolar charge interactions
predominantly due to the octahedron of fluorines, minus sign is
related to the negative charge at the F ions), B$_2^0$= +2 meV
(quadrupolar (axial term) charge interactions, positive sign
corresponds to the compression along z-axis); B$_2^2$= -7 meV
(quadrupolar (planar term) charge interactions, negative sign
corresponds to the elongation along y-axis). We take also into
account the intra-atomic spin-orbit coupling $\lambda _{s-o}$ of
-150 meV - its effect is not visible in the splitting of the
$^{2}E_{g}$ subterm (Fig. 1c) but the spin-orbit coupling affects
the eigenfunctions of the $^{2}E_{g}$ subterm and its magnetic
characteristics. As a consequence the moment of the ground-state
doublet is not any more the integer 1 $\mu_{B}$ but $\pm$1.26
$\mu_{B}$ for the present calculations (Fig. 1e). The detailed
eigenfunctions, the energy states, and the magnetic moment can be
calculated like we have demonstrated for many compounds (Fe$^{2+}$
ion in FeBr$_{2}$ \cite{9} and FeO \cite{10}, Co$^{3+}$ in
LaCoO$_{3}$ \cite{11}, Ti$^{3+}$ in YTiO$_{3}$ \cite{12},
Co$^{2+}$ in CoO or Ni$^{2+}$ in NiO \cite{13}) both in the
paramagnetic state and in the magnetically-ordered state.

The respective Hamiltonian is considered in the LS space that is
the 10 dimensional spin-orbital space $\left|
LSL_{z}S_{z}\right\rangle $. Despite of the relative weakness of
the s-o coupling for the $d$ ions in comparison to the strength
of the crystal-field interactions we have performed direct
calculations treating all terms in the Hamiltonian on the same
footing. Due to the spin-orbit coupling the involved functions
are not any more the pure cubic $e_{g}$ ($x^{2}-y^{2}$ or
3$z^{2}-r^{2}$) and $t_{2g}$ states.

An analysis of the effect of the sign of the tetragonal
off-octahedral distortion leads to a conclusion that for the
elongation along $z$-axis (c/a$>$1) the magnetic moments order
along the tetragonal axis. In case of the compression (c/a$<$1)
the moments are confined to the tetragonal plane. But then an
in-plane distortion has to occur in order to remove the in-plane
frustration - thanks this distortion a specific direction in the
plane can be selected. Again the moment is directed along the
most elongated bond - exactly as it is in case of
Cs$_{2}$AgF$_{4}$, where the ordered moment lies along the $y$
direction having the biggest length. The B$_{2}^{2}$ parameter is
related to the bond difference in the $a-b$ plane.

The calculated charge-formed ground-state has $\left\langle
S_{z}\right\rangle$ = $\pm $1.00, $\left\langle
L_{z}\right\rangle$ = $\pm $ 0.26 and the quadrupolar moment Q
=+5.97. The moments $m_{y}$ =$\pm $1.26 $\mu_{B}$ cancel each
other in the paramagnetic state, Fig. 2a. They reveal themselves
in the magnetic state when the Kramers-doublet ground-state
function becomes polarized because a molecular field is
self-consistently settled down.

Below T$_{c}$ there opens, as is seen in Fig. 2a, a spin-like gap
that amounts at T = 0 K to 2.6 meV. The spin-like gap is
associated with the splitting of the Kramers doublet ground state
in the ferromagnetic state. The magnetic ground state $\psi
_{GS+}$ has $\left\langle S_{z}\right\rangle$ = 1.00,
$\left\langle L_{z}\right\rangle$ = +0.26 and the resultant
moment of 1.26 $\mu_{B}$. The appearance of the magnetic state is
calculated self-consistently. It appears at the instability
temperature (T$_{c}$) in the temperature dependence of the CEF
paramagnetic susceptibility when
\begin{center}
\vspace {-0.1cm} $\chi_{CF}^{-1}(T_{c})$ = $n_{d-d} $
\end{center}
where $n_{d-d}$ is the molecular-field coefficient accounting
spin-dependent interactions. The ordering temperature of 14.9 K
yields $n_{d-d}$ = 13.8 T/$\mu_{B}$ (= 9.27 K/$\mu_{B}^{2}$) for
all magnetic interactions of the given Ag-moment with its magnetic
neighbours. The Ag moment experiences at T = 0 K a field of 17.4
T.

From the calculated free energy $F(T)$ we calculate all
thermodynamics like temperature dependence of the magnetic
moment, of the additional heat capacity $c_{d}$, of the
paramagnetic susceptibility $\chi_d$, of the 4$d$-shell quadrupolar moment
and many other properties similarly to those performed for
FeBr$_{2}$ \cite{9}, CoO \cite{13} and YTiO$_{3}$ \cite{12}.

Comparing our results with those of the Wu-Khomski's approach 1)
we claim that the distortion is not necessary for
Cs$_{2}$AgF$_{4}$ to be insulator - it is unphysical to think
that so small effect can produce so drastic change of electrical
properties, 2) we question an understanding that the insulating
gap occurs within {\bf the spin-polarized} $x^{2}-z^{2}$ state -
it would mean that Cs$_{2}$AgF$_{4}$ would be insulating only in
the magnetically-ordered state - although we do not have
experimental results at hands we believe that Cs$_{2}$AgF$_{4}$
is insulating both in the ferromagnetic state as well in the
paramagnetic state above 14.9 K, 3) the spin splitting of 2.6 meV
is more physically realistic for a compound with $T_{c}$ of 14.9
K than 1.5 eV \cite{8} corresponding to the thermal energy of
17400 K.

\section {Conclusions}

We have calculated consistently a value of the magnetic moment of
1.26 $\mu_{B}$ in Cs$_{2}$AgF$_{4}$ and its direction, along the
$y$ axis in the orthorhombic structure originating from the
tetragonal K$_{2}$NiF$_{4}$-type structure. We have derived the
spin and orbital moment and found that the orbital magnetic
moment cannot be ignored in any discussion of magnetic and
electronic properties of any $d$-containing compound. It confirms
the importance of the spin-orbit coupling which entangles the
spin and orbital degree of freedom. We have derived the
electronic structure both in the paramagnetic and ferromagnetic
state showing that this structure is only slightly modified by
the formation of the magnetic state. The respective
spin-polarization energy is only 1.3 meV/f.u. - it is almost 1000
times less than regarded in the recent theoretical papers. We
derive the strength of the octupolar charge interactions at the
Ag site to be 2.16 eV - it determines the $t_{2g}$-$e_{g}$
promotion energy (=10$Dq$). This $d-d$ excitation is a reason for
the lilac colour \cite{1} of Cs$_{2}$AgF$_{4}$. In our
understanding the insulating gap is much larger than 2.16 eV, say 4-5 eV, and
Cs$_{2}$AgF$_{4}$ is insulating both in the tetragonal and the
orthorhombic phase. Our results are important because the
$t_{2g}$-$e_{g}$ and spin-polarization energy are the basic
ingredient for any theory dealing with $d$ electron systems.
These studies prove that magnetic properties of Cs$_{2}$AgF$_{4}$
are predominantly determined by the atomic-scale lattice
distortions, crystal-field and the spin-orbit coupling of the
Ag$^{2+}$ ions, whereas charge fluctuations are of the minor
importance. An interplay of the spin-orbit coupling, lattice
distortions and the magnetic order is very subtle and involves
rather small energies, smaller than 5 meV making theoretical
studies difficult. We point out that all discussed by us
parameters are physical measurable parameters. Our approach can
be experimentally verified by the measurement of the $g$ factor
and the energy absorption at 2.0-2.3 eV.


\begin{thebibliography}{9}
\bibitem{1} S. E McLain, D. A. Tennant, J. F. C. Turner, T. Barnes,
M. R. Dolgos, Th. Proffen, B. C. Sales, and R. I. Bewley, Nature
Mater. {\bf 5}, 561 (2006); arXiv:cond-mat/0509194.

\bibitem{2} R.-H.Odenthal, D. Praus, and R. Hoppe, Z. Anorg. Allg.
Chem. {\bf 407}, 144 (1974).

\bibitem{3} T. Lancaster, S. J. Blundell, P. J. Baker, W. Hayes,
S. R. Giblin, S. E. McLain, F. L. Pratt, Z. Salman, E. A. Jacobs,
J. F. C. Turner, and T. Barnes, Phys. Rev. B {\bf 75}, 220408(R)
(2007).

\bibitem{4} D. Kasinathan, A. B. Kyker, and D. J. Singh, Phys. Rev. B
{\bf 73}, 214420 (2006).

\bibitem{5} D. Dai, M. -H. Whangbo, J. Kohler, C. Hoch, and A.
Villesuzanne, Chem. Mater. {\bf 18}, 3281 (2006).

\bibitem{6} Er-Jun Kan, Lan-Feng Yuan, Jinlong Yang, and J. G. Hou, Phys.
Rev. B {\bf 76}, 024417 (2007).

\bibitem{7} Xianfeng Hao, Yuanhui Xu, Zhijian Wu, Defeng Zhou, Xiaojuan Liu, and Jian Meng,
Phys. Rev. B {\bf 76}, 054426 (2007).

\bibitem{8} Hua Wu and D. I. Khomskii, Phys. Rev. B {\bf 76}, 155115 (2007).

\bibitem{9} Z. Ropka, R. Michalski, and R. J. Radwanski, Phys. Rev. B {\bf 63}, 172404 (2001).

\bibitem{10} R. J. Radwanski and Z. Ropka, Acta Physica {\bf 4}, 1 (2007), www.actaphysica.eu.

\bibitem{11} Z. Ropka and R. J. Radwanski, Phys. Rev. B {\bf 67}, 172401
(2003).

\bibitem{12} R. J. Radwanski and Z. Ropka, Acta Physica {\bf 5}, 5
(2007), www.actaphysica.eu; arXiv:cond-mat/0601005.

\bibitem{13} R. J. Radwanski and Z. Ropka, Acta Physica {\bf 1}, 26 (2006), www.actaphysica.eu; arXiv:cond-mat/0606604.
\end{thebibliography}
\end{document}